\documentclass[10pt,journal,a4paper]{IEEEtran}
 \pdfoutput=1 
\usepackage{graphicx}
\usepackage{subfigure}
\usepackage{float}
\usepackage{caption}
\usepackage{cleveref}

\usepackage{cite}
\title{Leaky Wave Antenna by Spin Photonic Topological Insulators Using 30-degree rhombic unit cell}
\begin{document}

\captionsetup[figure]{labelfont={bf},labelformat={default},labelsep=period,name={Fig.}}
\author{Sayyed~Ahmad~Abtahi\textsuperscript{1},
        Mohsen~Maddahali\textsuperscript{2},
        and~Ahmad~Bakhtafrouz\textsuperscript{3}\\
        Department of Electrical and Computer Engineering \\
        Isfahan University of Technology,
        Isfahan 84156-83111, Iran\\
        \textsuperscript{1}abtahi.a@ec.iut.ac.ir; \textsuperscript{2}maddahali@iut.ac.ir; \textsuperscript{3}bakhtafrouz@iut.ac.ir}
\maketitle

\begin{abstract}
A new spin photonic topological insulator model with a 30-degree rhombic unit cell is presented and analyzed. Topological band gaps and the robustness against the defects are shown. The edge modes are studied via full-wave simulation of a ribbon of unit cells. By using the structure in the fast wave region, a leaky wave antenna is proposed with a gain of 18.2dB.
\end{abstract}


\IEEEpeerreviewmaketitle

\section{Introduction}

\IEEEPARstart{L}{eaky} wave antennas are one of the most practical antennas that are realized widely with metasurfaces, which have the advantage of being low profile and easy to manufacture\cite{ref1}. Operation of these antennas is based on the modes in the fast wave region of their dispersion diagram \cite{ref13}. Some of such antennas are designed to have multi beam \cite{ref14} or have a wide range of beam scanning\cite{ref15}.

In recent years, with the introduction of the Photonic Topological Insulator (PTIs) \cite{ref2,ref3,ref4,ref5} and realizing them with metasurfaces, a new model was created, which is called topological metasurfaces\cite{ref6}. These structures provide advantages such as immunity from backscattering in sharp turns and manufacturing defects\cite{ref7}. The propagating states (edge states) in such systems will only appear at the interface of trivial and non-trivial materials, and the bulk will be an insulator.

The system presented in \cite{ref7,ref8} is a spin topological metasurface that uses a hexagonal metal patch and frame, which supports TE and TM modes, respectively. When the space between the patch and complementary frame becomes thin enough, The TE and TM modes hybridize bi-anisotropically and create the pseudo-spins, which opens a band gap at the symmetry point of K/K’. By changing the hexagonal unit cell to a 60-degree rhombic unit cell in \cite{ref9}, EM duality does enforce the twofold degeneracy, and consequently, the bandgap will occurs. The advantages of the rhombic unit cell are the simplicity of designing practical devices and integration with classical systems. 

In this paper, a 30-degree rhombus topological cell is proposed, and its differences from a 60-degree model are shown. With the advantages it offers, a new feasible multi-beam leaky wave antenna has been designed with a maximum gain of 18.5 GHz and a bandwidth of 2GHz, providing a 30-degree scan for each beam.

\section{Dispersin Diagram}

\begin{figure}[t!]
\begin{center}
\subfigure[]{
\includegraphics[scale=0.23]{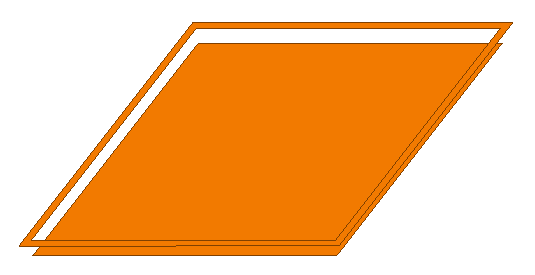}
}
\hfill
\subfigure[]{
\includegraphics[scale=0.23]{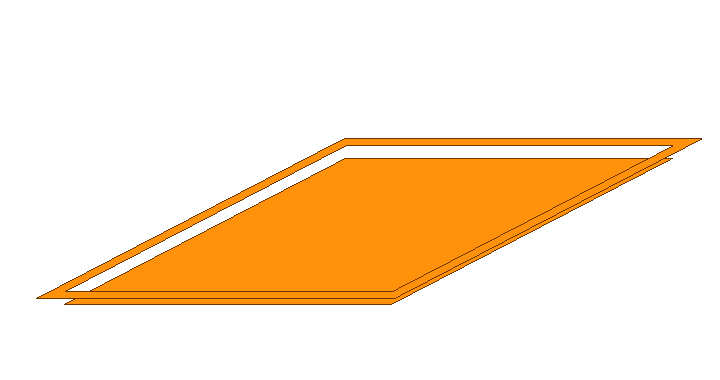}
}

\subfigure[]{
\includegraphics[height=3cm,width=4cm]{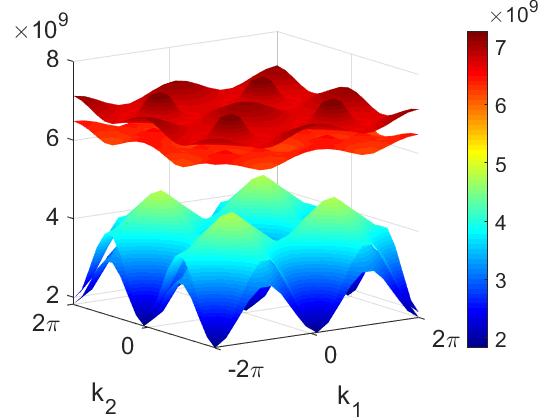}
}
\hfill
\subfigure[]{
\includegraphics[height=3cm,width=4cm]{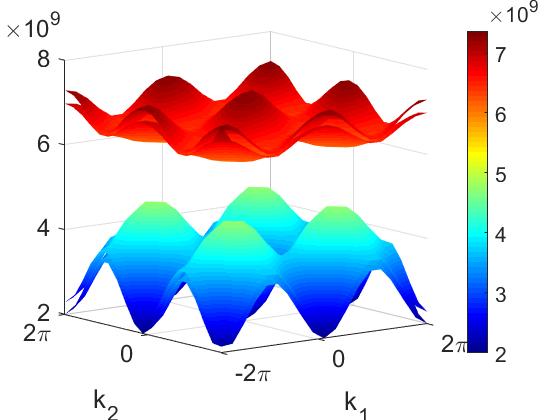}
}
\end{center}
\caption{comparison between 3D dispersion diagram of two rhombic unit cells with equal period of 20 mm and broader width of 0.433 mm. The not shown dielectric between the patch and frame is Rogers/Duroid 5880 with the thickness of 1.57 mm.}
\label{fig1}
\end{figure}

As the 3D dispersion diagram of the 30 degree rhombic unit cell shown in the Fig. \ref{fig1}, the width of the band gap becomes smaller than the similar 60 degree rhombic unit cell.

\begin{figure}[h!]
\begin{center}
\subfigure[]{
\includegraphics[scale=0.5]{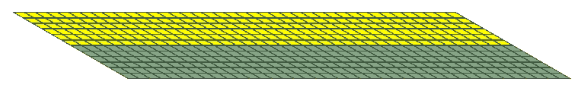}
}

\subfigure[]{
\includegraphics[scale=0.4]{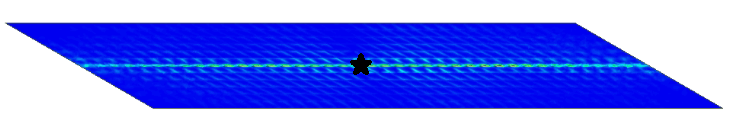}
}
\end{center}
\caption{The edge mode propagation at 5.7 GHz through the interface, excited by an Hertian $E_z$ dipole in the middle of the structure shown by the black star.}
\label{fig2}
\end{figure}

Here, as done in \cite{ref7,ref9}, two lattices of unit cells were placed beside each other, with the patches and frames flipped along z to observe the edge mode. In Fig. \ref{fig2}, the simulation results show that the structure is excited with a Hertzian dipole to study nontrivial edge mode propagation through the interface at 5.7 GHz.

\begin{figure}[h!]
\begin{center}
\subfigure[]{
\includegraphics[height=3cm,width=8cm]{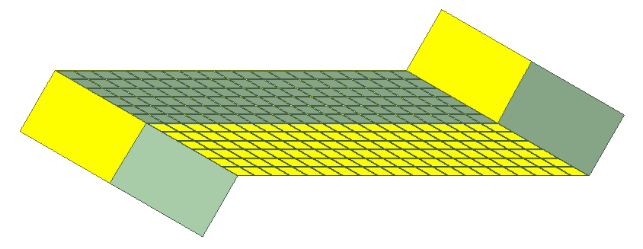}
}

\subfigure[]{
\includegraphics[height=5cm,width=9cm]{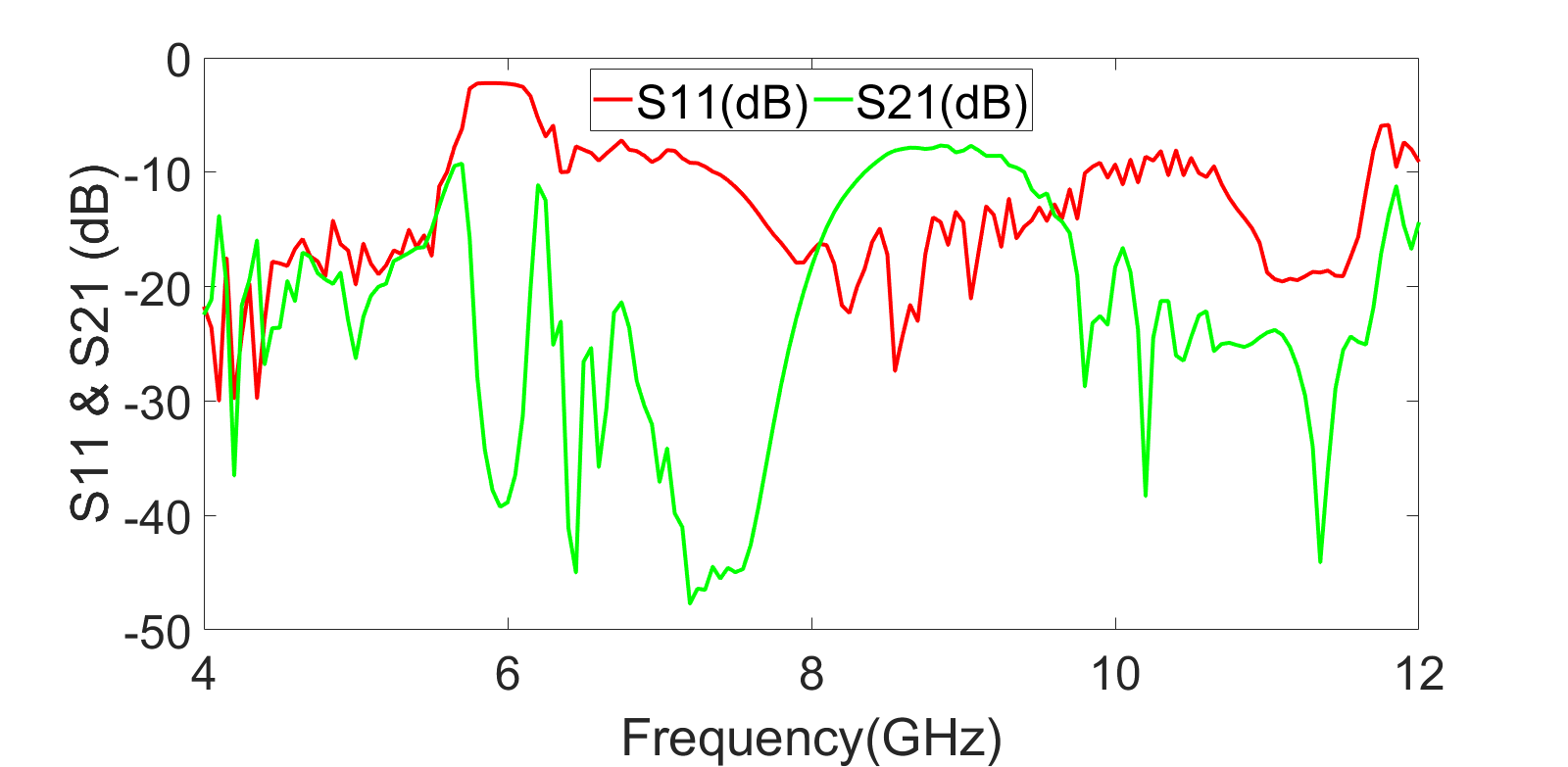}
}
\end{center}
\caption{Exciting the structure with ASL and the result}
\label{fig3}
\end{figure}

\begin{figure}[h!]
\begin{center}
\subfigure[]{
\includegraphics[height=3cm,width=8cm]{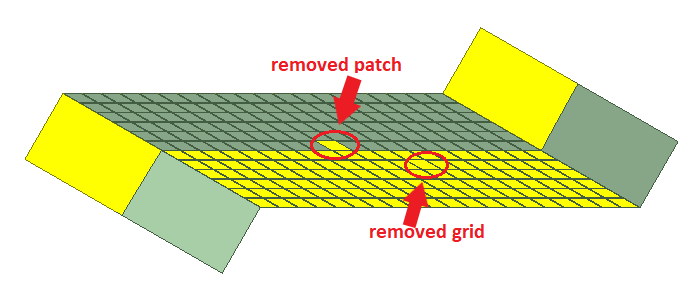}
\label{fig4a}
}

\subfigure[]{
\includegraphics[height=5cm,width=9cm]{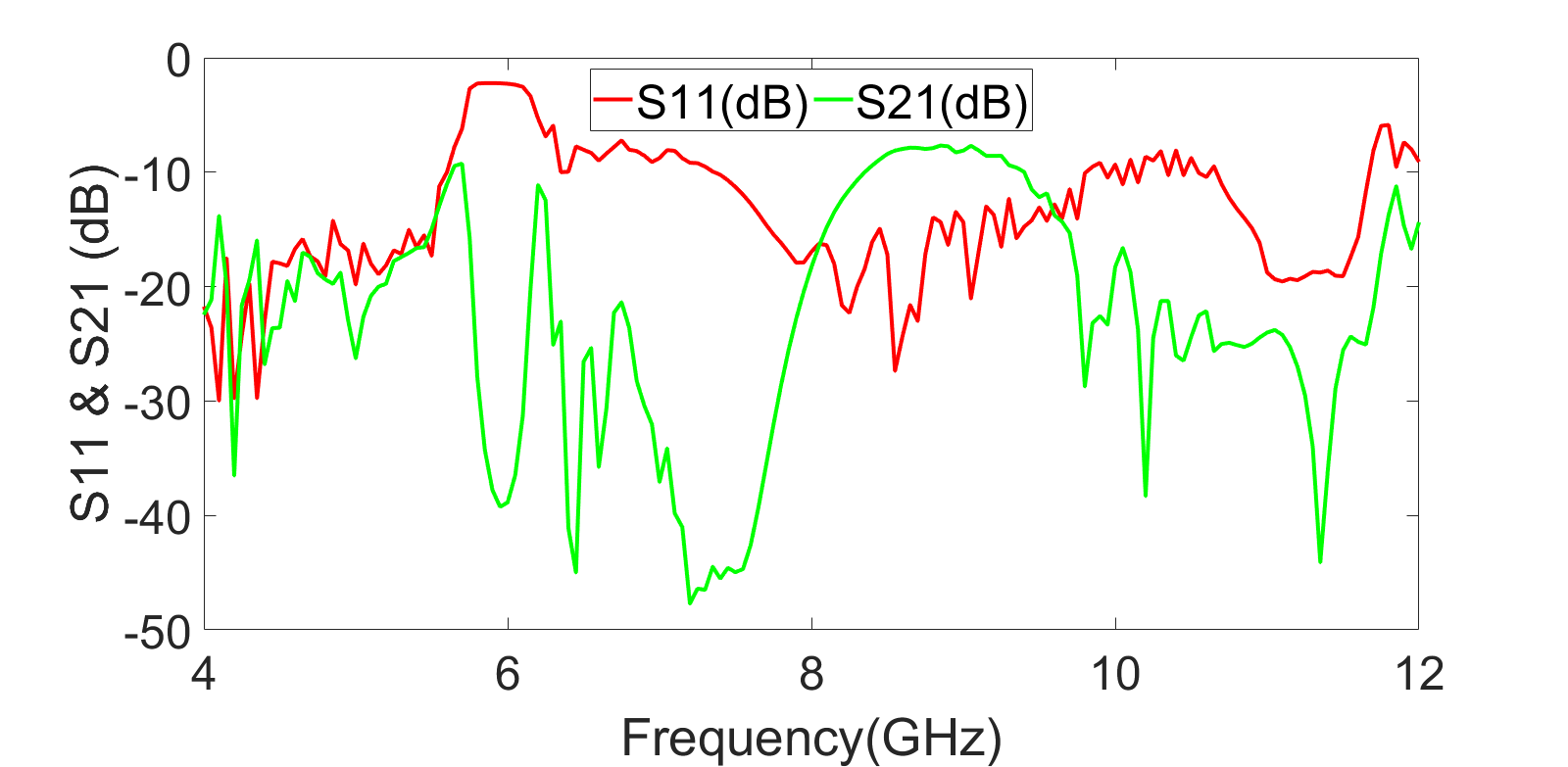}
\label{fig4b}
}
\end{center}
\caption{\subref{fig4a}The structure with defects and \subref{fig4b}the return and insertion loss in the presence of the defects}
\label{fig4}
\end{figure}

Similar for the hexagonal unit cell reported in \cite{ref10}, antipodal slot line (ASL) transition to a topological structure is used for excitation. By the way, because of the geometry of the unit cell, ASL is more adaptive to these structure than the hexagonal one.


Furthermore to observe the effect of defects, we apply some defects to the structure by omitting some patch and frames..The result in Fig. \ref{fig4} showing that structure with the defects almost has no effect on the propagation and the edge mode is robust against the defects.

\begin{figure}[t!]
\begin{center}
\subfigure[]{
\scalebox{1}[-1]{\includegraphics[height=0.85cm,width=9cm]{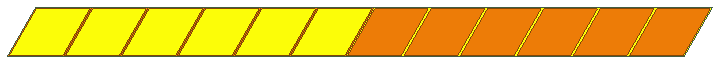}}
\label{fig5a}
}

\subfigure[]{
\includegraphics[height=1cm,width=9cm]{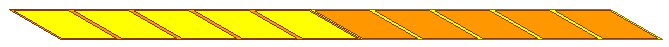}
\label{fig5b}
}
\end{center}
\caption{The super-cells for the simulation of the edge mode}
\label{fig5}
\end{figure}

To confirm the edge mode presence in the frequency band gaps, one line of the structure is separated as a super-cell and the eigenmodes are obtained.Fig. \ref{fig5} shows the super-cell's setup for the simulation by Ansys HFSS.

By comparison of the 30-degree and 60-degree rhombic unit cell, with equal dimensions, it is observed that for a 60-degree super-cell, the edge mode will not cover all of the bandgaps, but for the 30-degree one, the remained band gap is smaller.

\begin{figure}[h!]
\begin{center}
\subfigure[]{
\includegraphics[height=3.25cm,width=4cm]{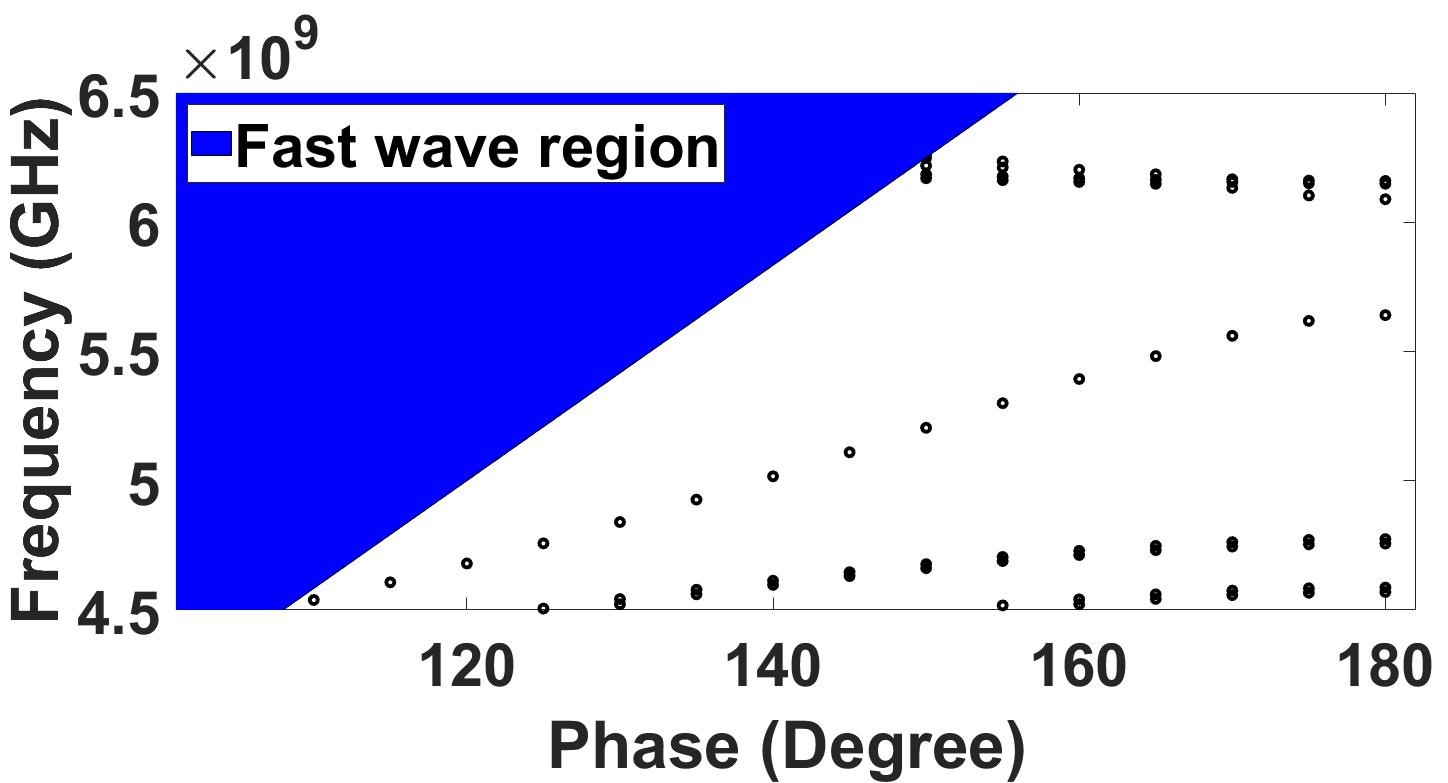}
\label{fig6a}
}
\hfill
\subfigure[]{
\includegraphics[ height=3.25cm,width=4cm ]{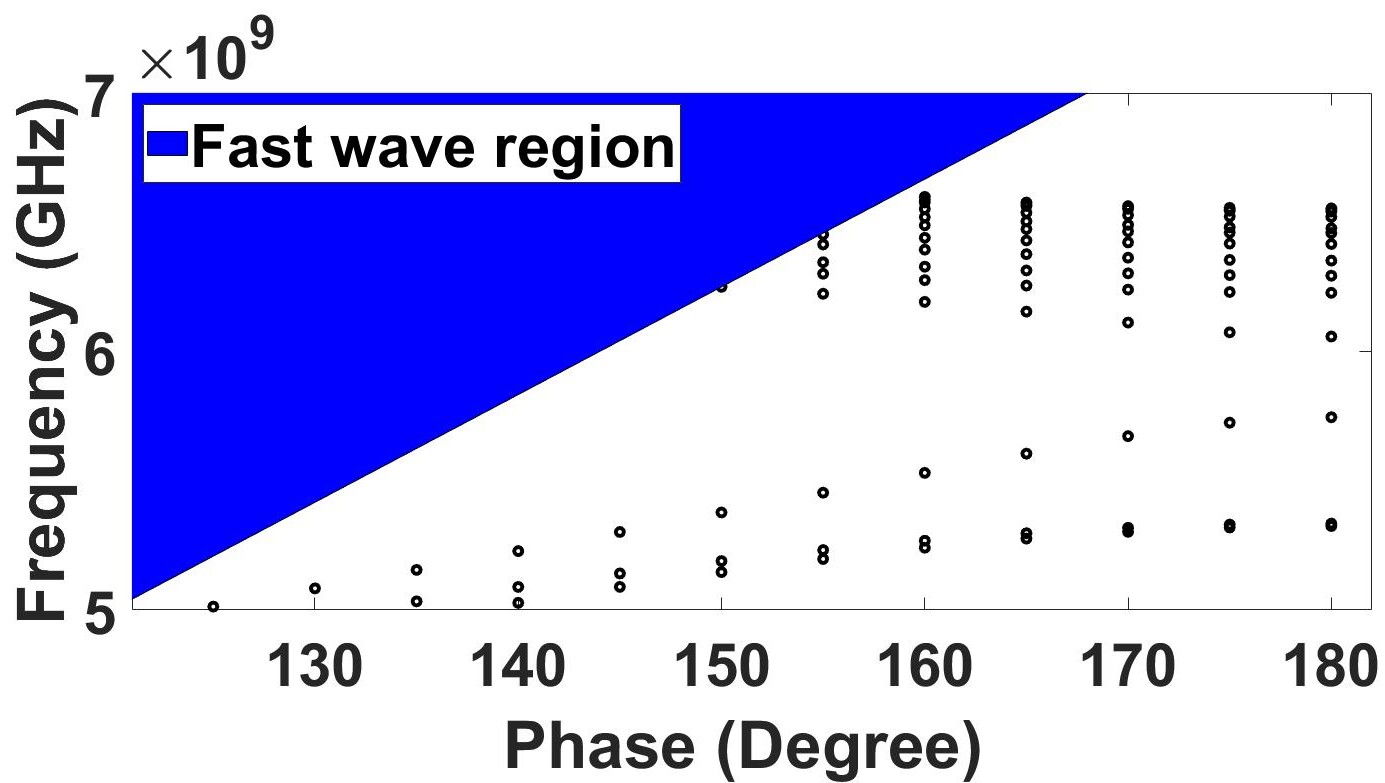}
\label{fig6b}
}
\end{center}
\caption{Dispersion diagram of supercell of \subref{fig6a} 60 degree rhombic cells and \subref{fig6b} 30 degree rhombic cells}
\label{fig6}
\end{figure}

\section{Antenna Application}

In \cite{ref11,ref12}, the Chern and spin PTIs were proposed for antenna application based on leaking out the energy. In \cite{ref11}, providing defects techniques are used to get the leakage in the guiding mode, and in \cite{ref12},the leakage is obtained by calculation of the dispersion diagram of the super-cell and finding the edge mode in the fast wave region. Both of them are not feasible and did not report their pattern characteristics.

\begin{figure}[t!]
\begin{center}
\includegraphics[height=4cm,width=9cm]{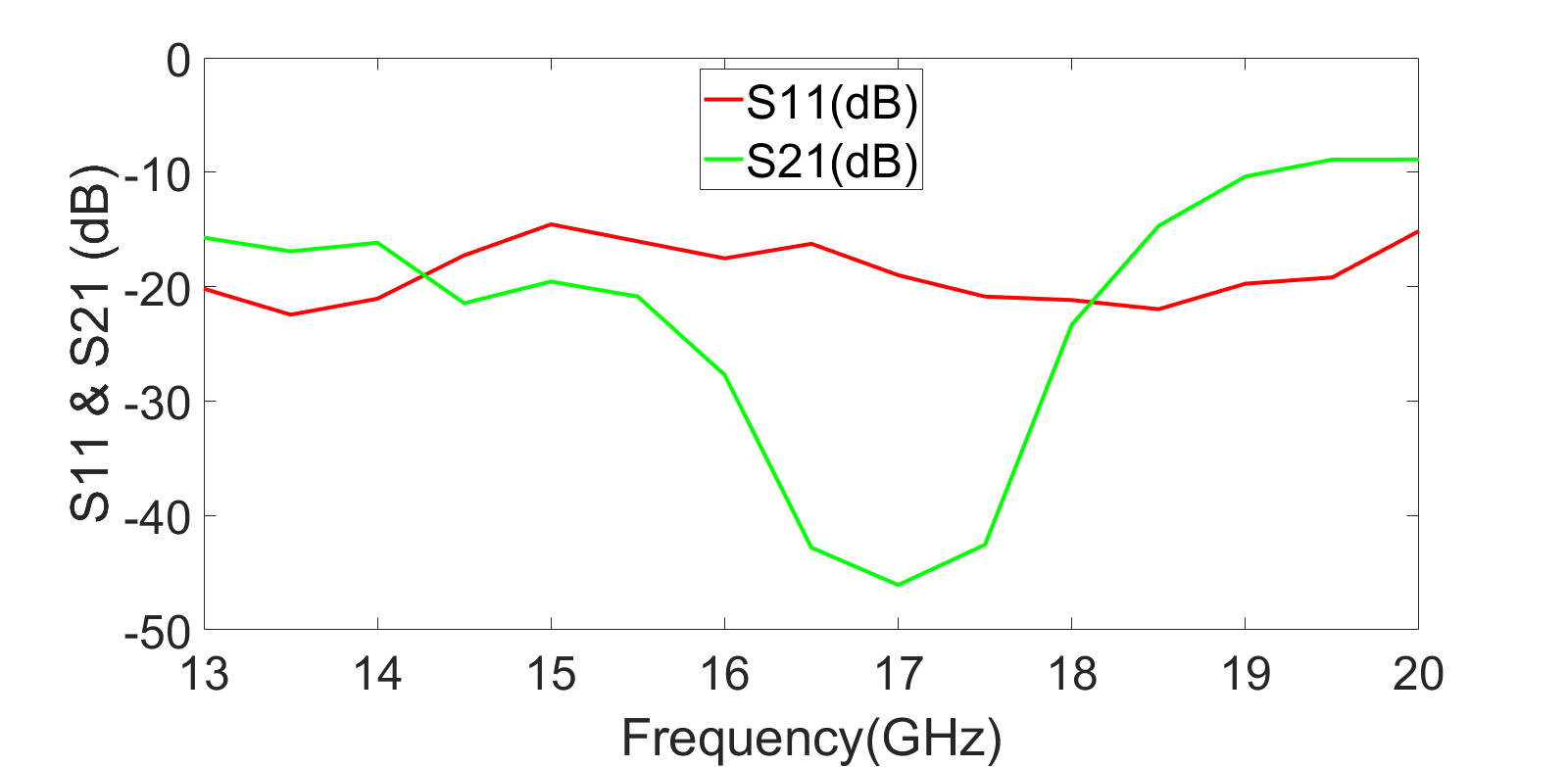}
\end{center}
\caption{S11 and S21 in the frequency range for operating as leaky wave.}
\label{fig7}
\end{figure}

In this paper, a leaky wave antenna based on a topological insulator is proposed, which uses a 30-degree rhombic complementary unit cell, on a 1.57-mm-thick Rogers/Duroid 5880 substrate, with a period of 10mm and border width of 0.5 mm. The design is based on \cite{ref12} to find the leaky modes for the antenna. So the upper band gaps used to operate in the leaky wave region, where both S11 and S21 are almost less than -10dB (Fig. \ref{fig7}). 

\begin{figure}[b!]
\begin{center}
\subfigure[]{
\includegraphics[height=6cm,width=9cm]{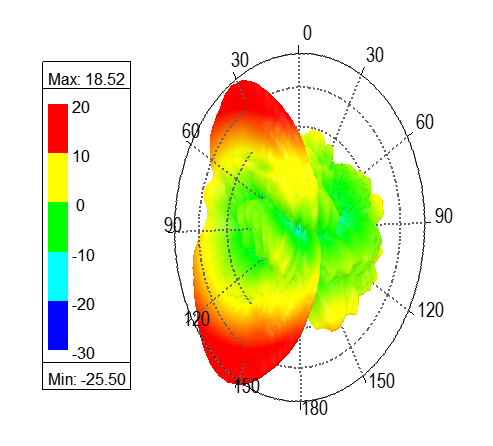}
\label{fig8a}
}

\subfigure[]{
\includegraphics[height=5cm,width=9cm]{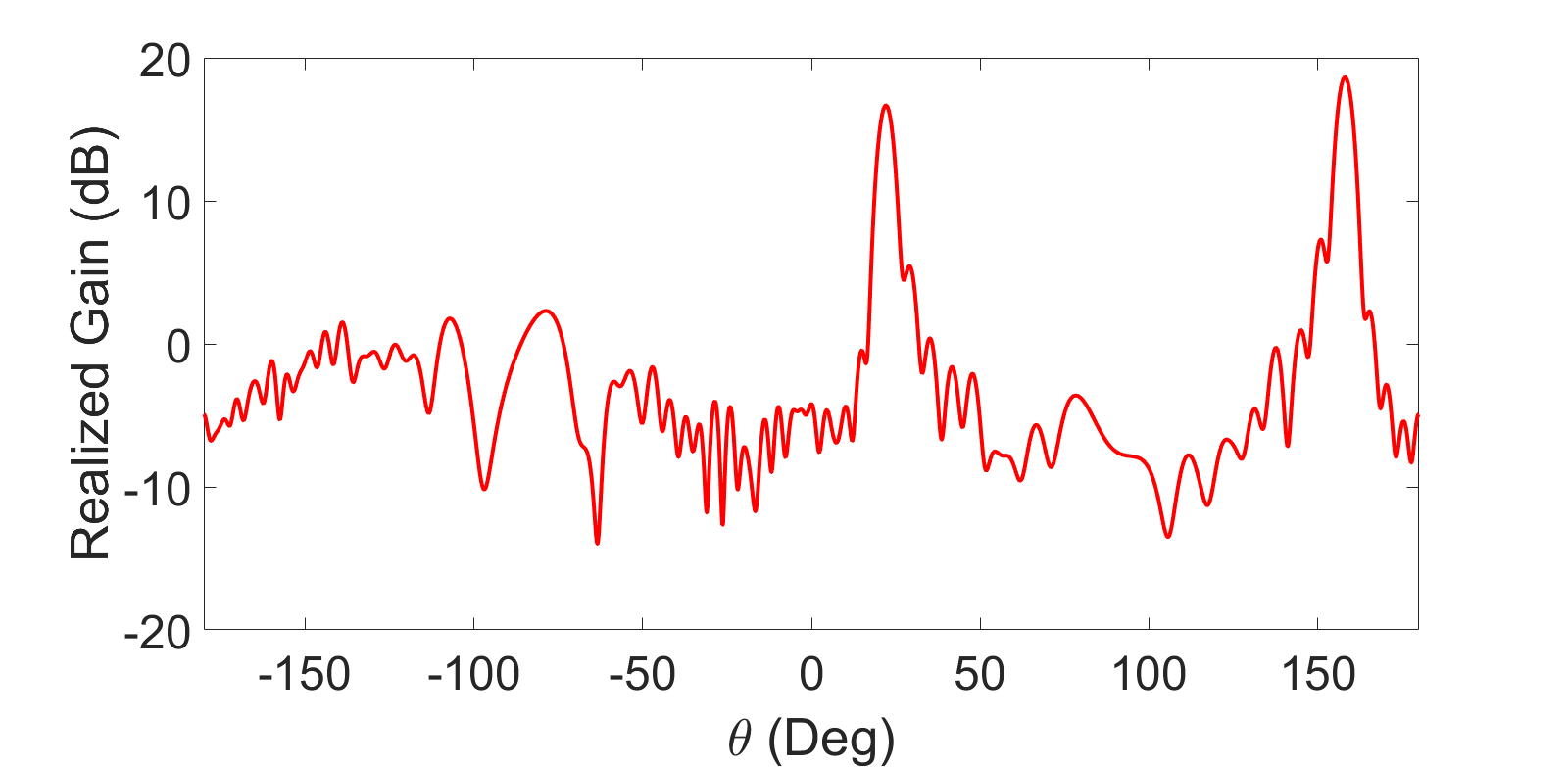}
\label{fig8b}
}
\end{center}
\caption{Realized Gain of the proposed antenna at 18.5 GHz.}
\label{fig8}
\end{figure}

\begin{figure}[h!]
\begin{center}
\includegraphics[height=4cm,width=10cm]{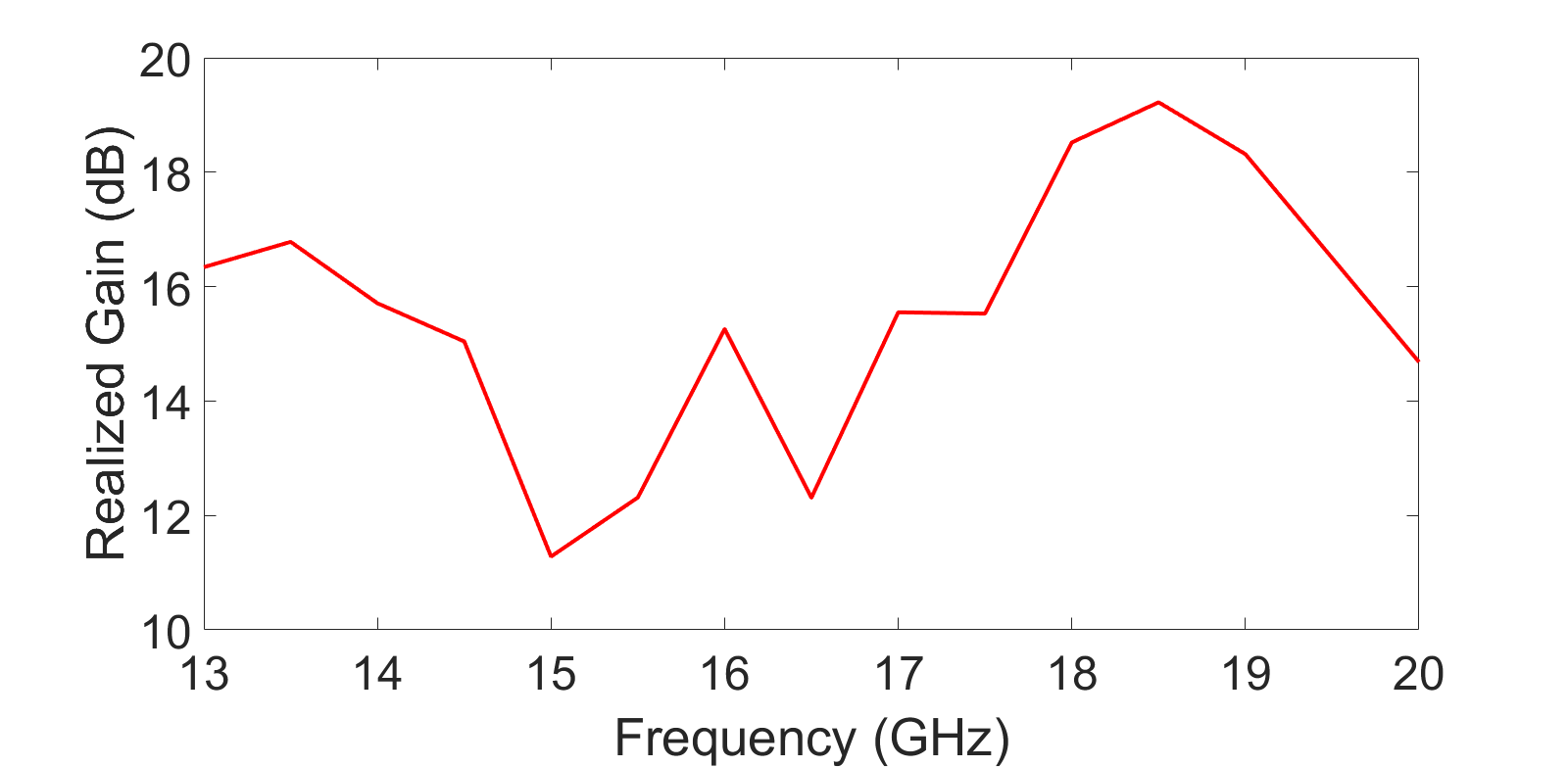}
\end{center}
\caption{Bandwidth of gain.}
\label{fig9}
\end{figure}

The antenna pattern, which is shown in Fig. \ref{fig8} has two beams with a maximum realized gain of 18.52dB at 18.5 GHz with the HPBW of is 3 degrees. As the radiation aperture is a rhombus with an area of 12000 (mm2), the aperture efficiency is nearly 32 percent. The bandwidth of gain is about 2 GHz around 18.5 GHz (Fig. \ref{fig9}).

\begin{figure}[t!]
\begin{center}
\includegraphics[height=10cm,width=10cm]{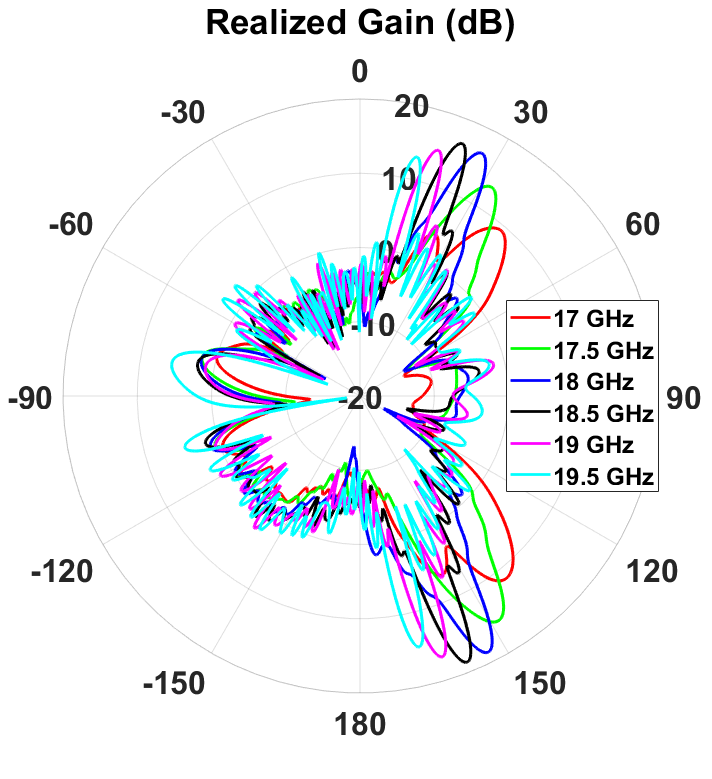}
\end{center}
\caption{Beam scanning by frequency changing between 17 to 19.5 GHz.}
\label{fig10}
\end{figure}

One of the most important applications of the leaky wave antenna is the ability of beam scanning. The proposed antenna can show about a 30-degree beam scan in the range of 9.5 to 40.9 and 141 to 171.5 degrees, by changing the frequency in the leaky mode region between 17 to 20 GHz (Fig. \ref{fig10}).

\section{Conclusion}

In this paper, a new 30-degree unit cell is proposed to demonstrate a topological microwave waveguide. We show the propagation of edge mode through the interface and the robustness against the defects with the transition of classical to topological metasurface for the structure made by rhombic unit cells. A leaky wave antenna was introduced based on the proposed topological metasurface by determining the operation frequency in the fast wave region.  The designed antenna is multibeam with the beam scan of 30 degrees and realized gain of 19.2dB at 18.5GHz.

\bibliographystyle{IEEEtran}
\bibliography{references1}

\begin{thebibliography}{10}
\providecommand{\url}[1]{#1}
\csname url@samestyle\endcsname
\providecommand{\newblock}{\relax}
\providecommand{\bibinfo}[2]{#2}
\providecommand{\BIBentrySTDinterwordspacing}{\spaceskip=0pt\relax}
\providecommand{\BIBentryALTinterwordstretchfactor}{4}
\providecommand{\BIBentryALTinterwordspacing}{\spaceskip=\fontdimen2\font plus
\BIBentryALTinterwordstretchfactor\fontdimen3\font minus
  \fontdimen4\font\relax}
\providecommand{\BIBforeignlanguage}[2]{{%
\expandafter\ifx\csname l@#1\endcsname\relax
\typeout{** WARNING: IEEEtran.bst: No hyphenation pattern has been}%
\typeout{** loaded for the language `#1'. Using the pattern for}%
\typeout{** the default language instead.}%
\else
\language=\csname l@#1\endcsname
\fi
#2}}
\providecommand{\BIBdecl}{\relax}
\BIBdecl

\bibitem{ref1}
D.~R. Jackson, C.~Caloz, and T.~Itoh, ``Leaky-wave antennas,''
  \emph{Proceedings of the IEEE}, vol. 100, no.~7, pp. 2194--2206, 2012.

\bibitem{ref13}
J.~L. Volakis, \emph{Antenna engineering handbook}.\hskip 1em plus 0.5em minus
  0.4em\relax McGraw-Hill Education, 2007.

\bibitem{ref14}
D.~Gonz{\'a}lez-Ovejero, G.~Minatti, G.~Chattopadhyay, and S.~Maci, ``Multibeam
  by metasurface antennas,'' \emph{IEEE Transactions on Antennas and
  Propagation}, vol.~65, no.~6, pp. 2923--2930, 2017.

\bibitem{ref15}
S.~Pavone, M.~Albani, S.~Maci, E.~Martini, and F.~Caminita, ``Beam scanning
  metasurface antennas,'' in \emph{2018 IEEE International Symposium on
  Antennas and Propagation \& USNC/URSI National Radio Science Meeting}.\hskip
  1em plus 0.5em minus 0.4em\relax IEEE, 2018, pp. 1443--1444.

\bibitem{ref2}
T.~Ozawa, H.~M. Price, A.~Amo, N.~Goldman, M.~Hafezi, L.~Lu, M.~C. Rechtsman,
  D.~Schuster, J.~Simon, O.~Zilberberg \emph{et~al.}, ``Topological
  photonics,'' \emph{Reviews of Modern Physics}, vol.~91, no.~1, p. 015006,
  2019.

\bibitem{ref3}
H.-X. Wang, G.-Y. Guo, and J.-H. Jiang, ``Band topology in classical waves:
  Wilson-loop approach to topological numbers and fragile topology,'' \emph{New
  Journal of Physics}, vol.~21, no.~9, p. 093029, 2019.

\bibitem{ref4}
S.~Singh, R.~J. Davis, J.~B. Dia’aaldin, J.~Lee, S.~M. Kandil, E.~Wen,
  X.~Yang, Y.~Zhou, P.~R. Bandaru, and D.~F. Sievenpiper, ``Advances in
  metasurfaces: Topology, chirality, patterning, and time modulation,''
  \emph{IEEE Antennas and Propagation Magazine}, vol.~64, no.~4, pp. 51--62,
  2021.

\bibitem{ref5}
R.~Davis, Y.~Zhou, P.~Bandaru, D.~Sievenpiper \emph{et~al.}, ``Photonic
  topological insulators: A beginner's introduction [electromagnetic
  perspectives],'' \emph{IEEE Antennas and Propagation Magazine}, vol.~63,
  no.~3, pp. 112--124, 2021.

\bibitem{ref6}
A.~Grbic and S.~Maci, ``Em metasurfaces [guest editorial],'' \emph{IEEE
  Antennas and Propagation Magazine}, vol.~64, no.~4, pp. 16--22, 2022.

\bibitem{ref7}
D.~J. Bisharat and D.~F. Sievenpiper, ``Electromagnetic-dual metasurfaces for
  topological states along a 1d interface,'' \emph{Laser \& Photonics Reviews},
  vol.~13, no.~10, p. 1900126, 2019.

\bibitem{ref8}
J.~B. Dia'aaldin and D.~F. Sievenpiper, ``Topological metasurfaces for
  symmetry-protected electromagnetic line waves,'' in \emph{Metamaterials,
  Metadevices, and Metasystems 2019}, vol. 11080.\hskip 1em plus 0.5em minus
  0.4em\relax SPIE, 2019, pp. 20--26.

\bibitem{ref9}
R.~J. Davis and D.~F. Sievenpiper, ``Robust microwave transport via nontrivial
  duality-based rhombic unit cells,'' in \emph{2021 IEEE International
  Symposium on Antennas and Propagation and USNC-URSI Radio Science Meeting
  (APS/URSI)}.\hskip 1em plus 0.5em minus 0.4em\relax IEEE, 2021, pp. 619--620.

\bibitem{ref10}
R.~J. Davis, D.~J. Bisharat, and D.~F. Sievenpiper, ``Classical-to-topological
  transmission line couplers,'' \emph{Applied Physics Letters}, vol. 118,
  no.~13, p. 131102, 2021.

\bibitem{ref11}
Y.~Lumer and N.~Engheta, ``Topological insulator antenna arrays,'' \emph{ACS
  Photonics}, vol.~7, no.~8, pp. 2244--2251, 2020.

\bibitem{ref12}
S.~Singh, D.~Bisharat, and D.~Sievenpiper, ``Topological antennas: Aperture
  radiators, leaky-wave surfaces, and orbital angular momentum beam
  generation,'' \emph{Journal of Applied Physics}, vol. 130, no.~2, p. 023101,
  2021.

\end{thebibliography}

\end{document}